# The lattice Boltzmann method for isothermal micro-gaseous flow and its application in shale gas flow: a review

Junjian Wang[a,b], Li Chen[c,b], Qinjun Kang[b], Sheik S Rahman[a,*]

[a]School of Petroleum Engineering, University of New South Wales, Sydney, NSW, Australia, 2033
[b]Earth and Environmental Sciences Division, Los Alamos National Laboratory, Los Alamos, NM, USA, 87545
[c]Key laboratory of Themo-Fluid Science and Engineering of MOE, School of Energy and Power Engineering, Xi'an Jiaotong University, Xi'an, Shanxi, China, 710049

**Abstract**

The lattice Boltzmann method (LBM) has experienced tremendous advances and been well accepted as a popular method of simulation of various fluid flow mechanisms on pore scale in tight formations. With the introduction of an effective relaxation time and slip boundary conditions, the LBM has been successfully extended to solve micro-gaseous related transport and phenomena. As gas flow in shale matrix is mostly in the slip flow and transition flow regimes, given the difficulties of experimental techniques to determine extremely low permeability, it appears that the computational methods especially the LBM can be an attractive choice for simulation of these micro-gaseous flows. In this paper an extensive overview on a number of relaxation time and boundary conditions used in LBM-like models for micro-gaseous flow are carried out and their advantages and disadvantages are discussed. Furthermore, potential application of the LBM in flow simulation in shale gas reservoirs on pore scale and representative elementary volume(REV) scale is evaluated and summarised. Our review indicates that the LBM is capable of capturing gas flow in continuum to slip flow regimes which cover significant proportion of the pores in shale gas reservoirs and identifies opportunities for future research.

## 1  Introduction

Shale gas reservoirs are thought to contain a significant proportion of hydrocarbon energy, and successful exploitation of such resource plays an increasingly important role in meeting world's demand for natural gas. Shale gas reservoirs are known to be fine grained sedimentary rocks which have complex porous structures with pores and fractures ranging from nano- to meso- scale[1][2], and in each level of pores and fractures different flow mechanisms are involved[3]. A better understanding of gas flow in shale is essential for future field development, and the prediction of permeability of gas has important applications in predicting the gas production rates from shale gas reservoir. Therefore, it is essential to develop accurate descriptive transport simulators which are capable of predicting flow dynamics in shale.

*Corresponding author at: School of Petroleum Engineering, University of New South Wales, Sydney, NSW,Australia,2033
Email address: sheik.rahman@unsw.edu.au (Sheik S Rahman)

Gas transport in shale is usually recognized by a dimensionless parameter, Knudsen number ($Kn$), which is the ratio of the gas mean free path to the characteristic length of the media. Current studies conclude that gas transport through shale matrix can best be characterized by $Kn$ in slip flow (0.001< $Kn$ <0.1) and transition flow (0.1< $Kn$ <10) regimes[4][5](see Fig.1). Under these conditions, continuum hypothesis is broken down and other rarefied gas transport mechanisms such as slip flow and Knudsen diffusion start to dominate the flow. Additionally, as a source rock, the presence of organic matter (kerogen) in shale matrix adds complexities in gas flow simulation and brings in new fluid transport qualities to shale gas reservoir. Gas transport in nano-pores inside the kerogen involves adsorption/desorption as well as surface diffusion due to strong molecular interactions between gas and kerogen.

A variety of experimental and numerical studies shows that rarefication effects influence the shale permeability measurements by increasing the apparent permeability values[6][7][8][9][10][11][12]. The effect of adsorption gas and the following surface diffusion on the permeability of shale, however, is not well understood and less widely explored. On one hand, studies confirmed that the multilayer adsorption can take place in organic pores because of the capillary condensation phenomenon[13][14], which will lead to a lower permeability in shale reservoirs[15]. On the other, surface diffusion is confirmed that can take account for 25% of total flux at low pressure[16]. Wu et al.[17] stated that when the pore size is less than 2 nm, the contribution of surface diffusion to total mass transfer can be as much as 92.95%.

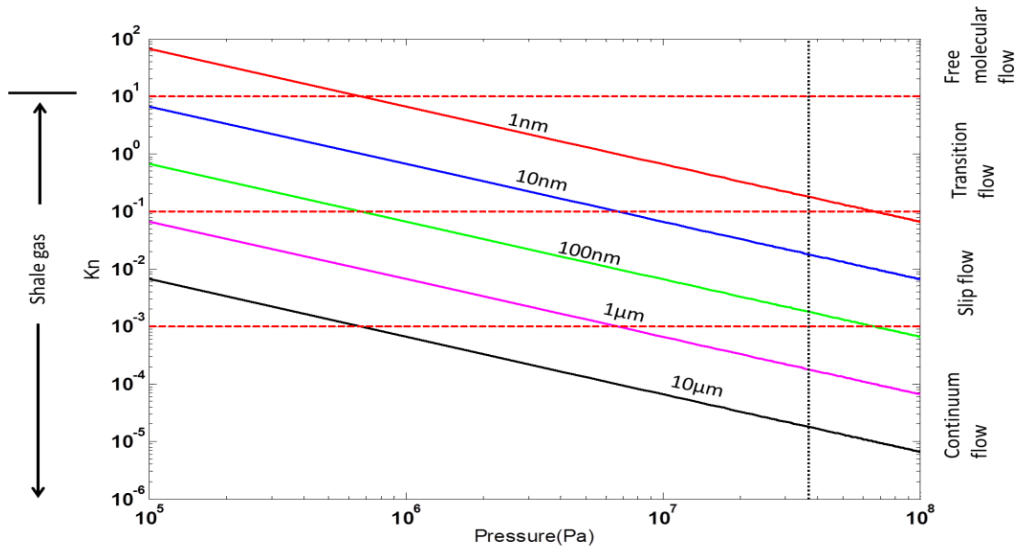

Figure 1: Knudsen number relationship to pore diameter and mean reservoir pressure at 400 K. Vertical dash line represent a typical reservoir pressure condition of 37 MPa.(Figure adapted from Javadapour et al.[3] and Sondergeld et al.[2] )

Generally, two possible mathematical approaches are proposed to properly describe the gas transport mechanism and to calculate gas apparent permeability of organic shale accomplished with the pore size distribution. The first approach is to modify the non-slip boundaries in continuum model by accounting for slip boundary conditions. Beskok-Karniadaki[18] derived a unified Hagen-Poiseuille-type formula to take account of all flow regimes. Later, Civan[19], Civan et, al.[20] and Florence et al.[21] proposed different forms of rarefaction coefficient for Beskok-Karniadaki model. By simply adding the mass transfer of adsorption gas into Beskok-Karniadaki model, the impact of the adsorbed gas and surface diffusion on gas apparent permeability are studied by Xiong et al.[22]. The second approach is the superposition of various transport mechanisms. Javadpour[3] combined slip flow and Knudsen diffusion into gas flux equation and derived an equation for apparent permeability. Freeman et al.[5] used dusty gas model to account for Knudsen diffusion in shale gas reservoir. Singh et al.[23] combined viscous flow with Knudsen diffusion in their non-empirical apparent permeability model(NAP), and the validation with previous experimental data indicates that the NAP can be used for Kn less than unity. Wu et al.[24] further proposed two weighted factors for viscous flow and Knudsen diffusion, respectively. The surface diffusion was also coupled in their apparent permeability model.

Most of the above mentioned previous studies are based on channel/tube flows and primary motivation of these studies is, to determine the apparent permeability of shale gas reservoirs. The complex geometry of shale matrix is modelled based on pore size distribution and tortuosity of connected pores and pore throats. Recently, well-established characterization techniques such as BIB-SEM, FESEM, FIB-SEM, TEM, micro-CT have been employed to identify a variety of pore structures in shale matrix[25][26][27][28][1][29]. This has allowed the development of image based porous structure that can be used to simulate gas flow. Among them, the Lattice Boltzmann method (LBM), which is vastly different from traditional CFD methods, has proven to be an effective flow simulation choice in porous media, as LBM algorithms are much easy to implement, especially in complex geometries and for multicomponent flows.

Historically, the LBM was derived from lattice gas automata (LGA)[30][31] and it was shown that LBE can also be directly derived by discretising the Boltzmann equation[32] [33]. Shortly after its introduction, the LBM became an attractive technique to study single/multi-phase flow[34][35][36][37][38] and reactive transport[39][40] in porous media, also covering groundwater flow[41], fabric materials[42] and fuel cells[43][44] etc. Detailed pore structure obtained by FIB-SEM and micro-CT have made the LBM a popular alternative to direct numerical solution of the Navier-Stokes equation for flows in tight rocks[45][46][47][48]. Implementation of the LBM is, however, far more than just provide a substitutable N-S solver on pore scale based on its intrinsically kinetic nature[49]. Advances in micro electrical mechanical system (MEMS) and nanotechnology have spurred interest in the use of the LBM for simulation of microfluidics and tremendous efforts have been made to advance the LBM since 2002. It is noteworthy that the LBM was initially extended for simulation of gaseous flows in slip flow regime[50-57]. Advances of the LBM have allowed us to simulate fluid flow in single capillary in transition flow regime[58-62]. In previous studies, most of the micro-gaseous flow was based on single relaxation time (SRT) model[51][54][55][56][58][63][64][65]. Luo[66] argued that slip velocity predicted by SRT is merely an artefact at the solid boundaries. For this reason, other LB models, such as two relaxation times (TRT)[67][68], multiple relaxation times (MRT)[59][61] and Filter-matrix lattice Boltzmann(FMLB)[60] were proposed. Results of these studies are in good agreement with that of the benchmark studies including force-driven Poiseuille flow, pressure driven Poiceuille flow, and planar Couette flow. For example, by incorporating the Bosanquet-type effective viscosity and applying slip boundary conditions, Li et al.[61] successfully used MRT models to simulate micro-channel gas flow at Kn of up to 3, this also gives a similar result as that of MRT model with a stops expression of effective viscosity proposed by Guo et al.[59]. Most Recently, Zhuo and Zhong[60] developed a Filter-matrix Boltzmann model with Bosanquet-type effective viscosity and produced reasonable results for micro-channel flow at Kn of up to 10.

Lattice Boltzmann method has been used to solve problems in petroleum industry for a decade, and several reviews on the LBM have been published[69-75]. These studies of microfluidic behaviours, however, have not been well addressed and a lot of debates exist in theory. The purpose of this paper is to summarise available typical boundary conditions and relaxation time used in the LBM-like algorithms for micro-gaseous flow, then simulation of fluid flow in shale gas reservoirs using representative LBM is discussed and finally feasibility and usefulness of the LBM are demonstrated.

## 2 Lattice Boltzmann method for isothermal micro-gaseous flow

In general, the lattice Boltzmann equation can be written as:

$$f_i(x + c_i \delta t, t + \delta t) - f_i(x, t) = \Omega_i(f_i) \quad (1)$$

Where $c_i$ indicates the finite number of discrete velocities of particles, $fi$ is the distribution function of particles with speed $c_i$, $\Omega_i$ is the collision term, $\delta x$ is the uniform lattice spacing and $\delta t$ is the time between two simulation iterations. The velocity models are illustrated in Fig. 2. The microscopic variables (density and velocity) are defined as:

$$\rho(x,t) = \sum_i f_i(x,t) \quad (2)$$

$$\rho(x,t)u(x,t) = \sum_i f_i(x,t)c_i \quad (3)$$

In order to account for the influence of body force, different schemes are described in literatures, these include but not limited to modifying velocity field with force term using Newton second law[76] and adding extra force term to the collision term[77][78]. Here we refer to the force term proposed by Guo et al.[77] as:

$$f_i(x + c_i \delta t, t + \delta t) - f_i(x, t) = \Omega_i(f_i) + \delta t F_i \quad (4)$$

Where, $F_i$ is the force term which is defined according to the collision operator and the velocity is adjusted as:

$$\rho(x,t)u^*(x,t) = \sum_i f_i(x,t)c_i + \frac{\delta t}{2} F \quad (5)$$

To solve the distribution function $f_i$ numerically, the lattice Boltzmann equation is further separated into a collision step and a propagation step as:

$$f_i^*(x,t) = f_i(x,t) + \Omega_i(f_i(x,t)) + \delta t F_i(x,t) \quad (6)$$

$$f_i(x + c_i \delta t, t + \delta t) = f_i^*(x,t) \quad (7)$$

This algorithm does not change the semantic of the equation but it allows the simulation to first compute the post collision distributions $f_i^*(x,t)$ in each fluid node $x$ at time $t$ and then to propagate them to the new destinations, which can be easily achieved by high performance parallel computing.

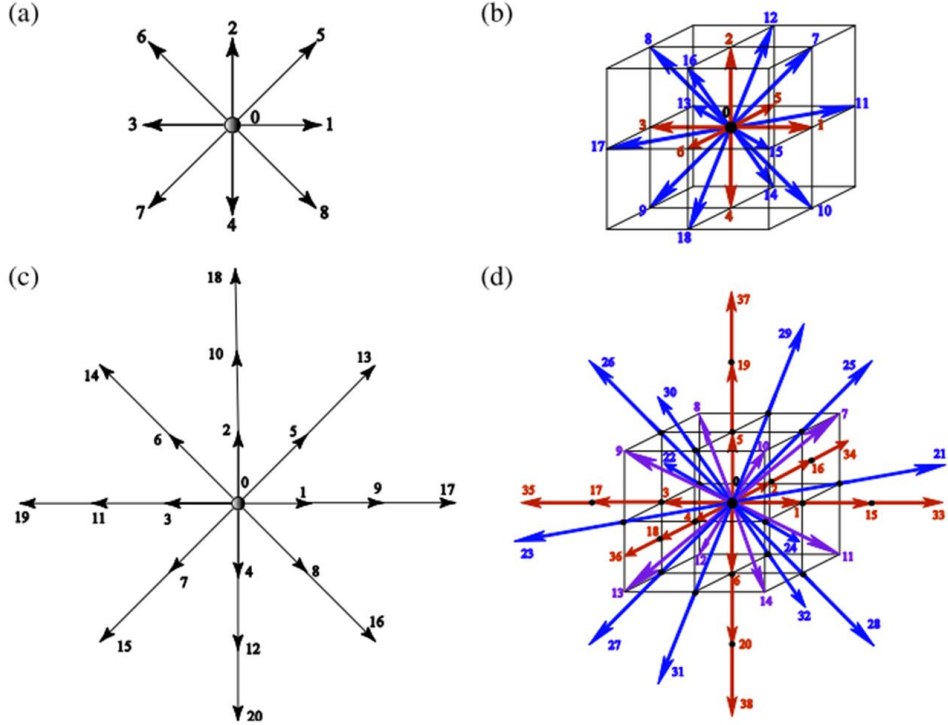

Figure 2: Discrete velocity models; (a)D2Q9 model, (b) D3Q19model, (c) D2Q21 model, (d) D3Q39 model

## 2.1 Collision operators

The collision operators describe the collision behaviour of particles at every lattice location, which represents the variation of distribution functions caused by collision between particles. The requirements are that they should be physically correct and efficiently computable. In the following, collision operators commonly used in micro-gaseous flow are presented.

### 2.1.1 Bhatnagar-Gross-Krook collision operator (BGK)

Bhatnagar, Gross and Krook[79] proposed an efficient simplification operator to approximate the collision term as:

$$\Omega_i(f_i) = -\tau^{-1}(f_i - f_i^{eq}) \tag{8}$$

Here, $\tau$ is the relaxation time and $f_i^{eq}$ is the Maxwell Boltzmann distribution of $c_i$ for a given macroscopic fluid velocity $u$ and density $\rho$. The Maxwell-Boltzmann distribution describes the density distribution of a fluid in its equilibrium state to which every fluid strives. The equilibrium distribution $f_i^{eq}$ in its discrete form can be expressed as:

$$f_i^{eq}(r,t) = \rho\omega_i[1 + \frac{3(c_i \cdot u)}{c^2} + \frac{9(c_i \cdot u)^2}{2c^4} - \frac{3(u \cdot u)}{2c^2}] \tag{9}$$

Suga[69] summarized the discrete velocities $c_i$, the sound speed $C_s$ and the weight parameter $\omega_i$ as per Fig. 2 (see Table 1).

Table 1: Parameters of the discrete velocity models for 2D/3D

| Models | $c_s^2/c^2$ | $c_i/c$ | $w_i$ |
|---|---|---|---|
| D2Q9 | 1/3 | (0,0) | $4/9 (i=0)$ |
| | | $(\pm 1,0),(0,\pm 1)$ | $1/9 (i=1-4)$ |
| | | $(\pm 1,\pm 1)$ | $1/36 (i=5-8)$ |
| D2Q21 | 2/3 | (0,0) | $91/324 (i=0)$ |
| | | $(\pm 1,0),(0,\pm 1)$ | $1/12 (i=1-4)$ |
| | | $(\pm 1,\pm 1)$ | $2/27 (i=5-8)$ |
| | | $(\pm 2,0),(0,\pm 2)$ | $7/360 (i=9-12)$ |
| | | $(\pm 2,\pm 2)$ | $1/432 (i=13-16)$ |
| | | $(\pm 3,0),(0,\pm 3)$ | $1/1620 (i=17-20)$ |
| D3Q19 | 1/3 | (0,0,0) | $12/36 (i=0)$ |
| | | $(\pm 1,0,0),(0,\pm 1,0),(0,0,\pm 1)$ | $2/36 (i=1-6)$ |
| | | $(\pm 1,\pm 1,0),(\pm 1,0,\pm 1),(0,\pm 1,\pm 1)$ | $1/36 (i=7-18)$ |
| D3Q39 | 2/3 | (0,0,0) | $1/12 (i=0)$ |
| | | $(\pm 1,0,0),(0,\pm 1,0),(0,0,\pm 1)$ | $1/12 (i=1-6)$ |
| | | $(\pm 1,\pm 1,\pm 1)$ | $1/27 (i=7-14)$ |
| | | $(\pm 2,0,0),(0,\pm 2,0),(0,0,\pm 2)$ | $2/135 (i=15-20)$ |
| | | $(\pm 2,\pm 2,0),(\pm 2,0,\pm 2),(0,\pm 2,\pm 2)$ | $1/432 (i=21-32)$ |
| | | $(\pm 3,0,0),(0,\pm 3,0),(0,0,\pm 3)$ | $1/1620 (i=33-38)$ |

The forcing term, $F_i$ can be specified with respect to the relaxation parameter $\tau$ and the body force $F$ as[77]:

$$F_i = (1 - \frac{1}{2\tau})\omega_i [\frac{c_i - u}{c_s^2} + \frac{c_i \cdot u}{c_s^4} c_i] \cdot F \tag{10}$$

Also in order to be valid, the LB equation requires the parameter to fulfil the following relation:

$$\mu = \rho c_s^2 (\tau - \frac{1}{2})\delta t \tag{11}$$

Where, $\mu$ being the dynamic viscosity of the fluid.

### 2.1.2 Two relaxation time collision operator (TRT)

The scheme for two relaxation time was developed by Ginzburg[80], for which the collision operator is split into symmetric and anti-symmetric parts as:

$$\omega(f_i) = -\tau_s^{-1}(f_i^s - f_i^{seq}) - -\tau_a^{-1}(f_i^a - f_i^{aeq}) \tag{12}$$

Here, relaxation time, $\tau_s$ related to shear viscosity, and relaxation time, $\tau_a$ related to energy fluxes, the symmetric and anti-symmetric components of distribution function and equilibrium distribution function can be computed as:

$$f_i^s = \frac{f_i + f_{\bar{i}}}{2}, f_i^{seq} = \frac{f_i^{eq} + f_{\bar{i}}^{eq}}{2}, \tag{13}$$

$$f_i^s = \frac{f_i - f_{\bar{i}}}{2}, f_i^{aeq} = \frac{f_i^{eq} - f_{\bar{i}}^{eq}}{2}, \tag{14}$$

### 2.1.3 Multiple Relaxation Times collision operator (MRT)

The lattice Boltzmann equation with multiple relaxation times (MRT) is a refined version of the lattice Boltzmann equation with the BGK collision operator. Instead of simulating the collision process by relaxing the distribution functions directly, the MRT collision operator relaxes the kinetic moments, which can be retrieved from the distribution functions. Relaxing the moments separately has the advantage that physical effects which cause the relaxation of distinct moments occurring on different time scales can easily be accommodated by determining the corresponding relaxation values. In addition, it is able to overcome the deficiencies of the BGK collision operator because it has more degrees of freedom, which can also be used to increase the numerical stability of the method significantly. The MRT collision operator is defined as follows[81]:

$$\Omega(f) = -(M^{-1}SM)(f - f^{eq}) \tag{15}$$

In D2Q9 model, $f = (f_0, f_1, \ldots, f_7, f_8)^T$ denotes the column vector of the distribution functions. $S$ is the non-negative relaxation matrix:

$$S = diag(\tau_\rho, \tau_e, \tau_\varepsilon, \tau_j, \tau_q, \tau_j, \tau_q, \tau_s, \tau_s)^{-1} \tag{16}$$

$M$ is an orthogonal transform matrix, which maps the distribution functions onto the moment space, and defined as:

$$M = \begin{bmatrix} 1 & 1 & 1 & 1 & 1 & 1 & 1 & 1 & 1 \\ -4 & -1 & -1 & -1 & -1 & 2 & 2 & 2 & 2 \\ 4 & -2 & -2 & -2 & -2 & 1 & 1 & 1 & 1 \\ 0 & 1 & 0 & -1 & 0 & 1 & -1 & -1 & 1 \\ 0 & -2 & 0 & 2 & 0 & 1 & -1 & -1 & 1 \\ 0 & 0 & 1 & 0 & -1 & 1 & 1 & -1 & -1 \\ 0 & 0 & -2 & 0 & 2 & 1 & 1 & -1 & -1 \\ 0 & 1 & -1 & 1 & -1 & 0 & 0 & 0 & 0 \\ 0 & 0 & 0 & 0 & 0 & 1 & -1 & 1 & 1 \end{bmatrix} \tag{17}$$

The distribution function and equilibrium function can be projected onto the moment space by using the transform matrix:

$$m = Mf = (\rho, e, \varepsilon, j_x, q_x, j_y, q_y, p_{xx}, p_{xy})^T \tag{18}$$

$$\begin{aligned} m_{eq} &= Mf_{eq} = (\rho, e^{eq}, \varepsilon^{eq}, j_x, q - x^{eq}, j_y, q_y^{eq}, p_{xx}^{eq}, p_{xy}^{eq})^T \\ &= \rho(1, -2 + 3|u|^2, 1 - 3|u|^2, u, -v, u, -v, u^2 - v^2, uv)^T \end{aligned} \tag{19}$$

All of these moments have a physically meaningful interpretation: $\rho$ is the density; $e$ is the energy mode; $\varepsilon$ is related to the energy square; $j_x$ and $j_y$ are the $x$ and $y$ components of the momentum; $q_x$ and $q_y$ correspond to the $x$ and $y$ components of the energy flux, $p_{xx}$ and $p_{xy}$ and are related to the diagonal and off-diagonal component of the stress tensor. $u$ and $v$ are $x$ and $y$ components of velocity $u$.

The forcing terms, $F_i$ can also be mapped onto the moment space which gives the vector $F$ as:

$$F = (0, 6u \cdot F, -6u \cdot F, F_x, -F_x, F_y, -F_y, 2(uF - vF_y), (uF + vF_y))^T \tag{20}$$

The general collision process can now be projected onto the moment space:

$$\begin{aligned} f^{\text{å}} &= f + \Omega f + \delta tF = M^{-1}Mf - M^{-1}SM(f - f^{eq}) + M^{-1}\delta tMF \\ &= M^{-1}(m - S(m - m^{eq}) + \delta t(I - \tfrac{S}{2})F) \end{aligned} \tag{21}$$

Then the dynamic viscosity, $\mu$ and the bulk viscosity of the fluid, $\zeta$ are given by:

$$\mu = \rho c_s^2(\tau_s - \tfrac{1}{2})\delta t \tag{22}$$

$$\zeta = \tfrac{1}{2}c_s^2(\tau_e - \tfrac{1}{2})\delta t \qquad (23)$$

It is worth to mention that the MRT-LBM can be reduced to the BGK-LBM if $S$ is set to $S_\alpha = 1/\tau \forall \alpha$. And by setting the relaxation rates for even-order non-conserved moments to $1/\tau$ and odder-order moments (i.e. $q_x$ and $q_y$) to:

$$\tau_q = 8\frac{2-\tau_s}{8-\tau_s} = 8\frac{2\tau-1}{8\tau-1} \qquad (24)$$

The MRT becomes TRT[82].

## 2.2 Capture gas-solid interfacial slip

In traditional fluid mechanics, the assumption of non-slip at a solid boundary is used as the boundary condition. The non-slip assumption, however, is broken down at micro- and nanoscales as many researchers have investigated the fluid/gas phenomenon[83][84]. Generally, the interfacial slip is generated by hydrophobicity in fluid flow and by Knudsen effect for gas flow[85]. To capture the fluid-solid interfacial slip, the Shan-Chen interparticle potential model is usually used, where the solid-fluid interaction is modelled via an explicit solid-fluid intermolecular potential to predict behaviours at the interface[85][86][87]. This approach provides better microscopic realism, however, it may face with more restrictive constraints in actual practice[52] and no definitive results have been provided to demonstrative its ability to capture gas slip in transition flow regime[88].

To capture the gas slip at the solid boundary, single phase LB model is widely used, and the slip boundary condition is adopted to implicitly consider the solid-fluid interaction. Tremendous efforts have been devoted to develop accurate and efficient boundary schemes for the LBM. In the following section we summarize typical boundary conditions that could be useful for micro-gas simulations.

### 2.2.1 Bounce back boundary condition(BB)

The half way bounce-back scheme[89] is typically utilized for its simplicity and its second-order accuracy. This boundary condition assumes that a particle which collides with the wall is reflected in opposite direction, which means that its momentum is reversed. In the implementation, particles leaving a boundary fluid node $x$ bounce back from the boundary to the original site in the reversed lattice velocity, this behaviour can be described by Eq. 25 as:

$$f_i^{bb}(x, t + \delta t) = f_{\bar{i}}^{å}(x, t) \qquad (25)$$

Where $f_{\bar{i}}^{å}$ denotes the opposing distribution function to $f_i$ leaving $x$ after collision at time $t$ such that $c_i = -c_{\bar{i}}$. The bounce-back condition was firstly used by Nie et al.[90] to mimic the microscopic flow, however, lately it has been shown that the boundary slip observed from Nie et al.[90] with a pure bounce-back scheme is just a numerical artefact[32] and cannot reflect the physical slip over the surface.

### 2.2.2 Specular reflection boundary condition(SR)

The specular reflection boundary condition is motivated by the observation of elastic collisions between a relatively light particle and a heavy boundary, the physical effect of such a collision is that the velocity component, which is orthogonal to the boundary, is reversed. This is described by the following equation for the D2Q9 model:

$$f_i^{sr}(x) = f_{i\prime}^{å}(x) \tag{26}$$

Where, $f_{i\prime}^{å}$ and $f_i^{sr}$ are approaching and specular reflecting distribution functions as shown in Fig. 3. This boundary condition is firstly applied by Lim et al.[91] to investigate pressure driven and shear driven micro-channel flows, however, it was found that the mesh size has a significant effect on the numerical results in their work[91][92] and pure specular reflection may overestimates the slip velocity[93].

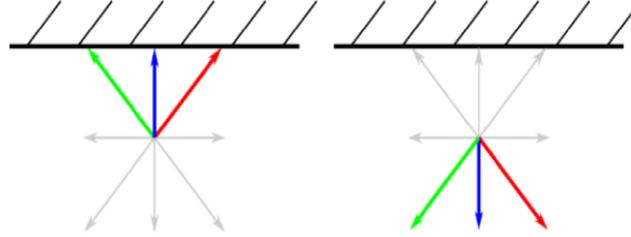

Figure 3: Illustration of the reflection process (in the left image, the distributions before the collision are highlighted. In the right image, the specular reflected distributions after collision with the wall are shown)[94]

### 2.2.3 Maxwellian diffuse reflection boundary condition(MD)

The Maxwellian diffusive reflection boundary condition is derived from the continuum kinetic theory for non-absorbing walls. The underlying idea of this boundary condition is that impinging particles lose the memory of their movement direction and are scattered back following a Maxwellian distribution in which the wall density, $\rho_w$ and the wall velocity, $u_w$ are known. The fully diffusive boundary condition can be described by the following equation[95]:

$$f_i^{dr}(x) = \frac{\sum_{(c_k-u_w)\cdot n<0} |(c_k-u_w)\cdot n| f_k^{å}(x)}{\sum_{(c_k-u_w)\cdot n>0} |(c_k-u_w)\cdot n| f_k^{eq}(\rho_w,u_w)} \cdot f_i^{eq}(\rho_w, u_w) \tag{27}$$

With $(c_k - u_w) \cdot n > 0$. The condition $(c_k - u_w) \cdot n < 0$ enforces that all incidental distributions are summed up and then redistributed over the outgoing distributions such that they obey the equilibrium distribution with mass conservation. The process of the boundary treatment is illustrated in Fig. 4.

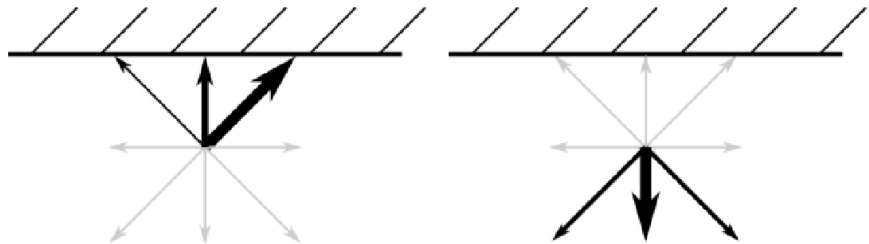

Figure 4: Illustration of the diffuse process(in the left image, the distributions before the collision and in the right image, the diffusive distributions collided with the wall are shown. The thickness of the arrows indicates the value of the distributions)[94]

This boundary condition has been implemented into LB model and the the results show good agreement with the analytical solution of Boltzmann equation for Kramer's problem as the $Kn$ tends to be zero[95], the similar concepts are used to simulate micro-gaseous flow with higher $Kn$ with D2Q9 LBM[51] and high order LBM[96]. Later Chai et al.[97] argued that the slip velocity is orverpredicted when MD scheme is applied to Poiseuille flow in a micro channel.

### 2.2.4 Combined form

The above mentioned three boundary conditions normally are not directly applied in the LB method, some improved version of the boundary conditions are proposed recently to mimic the macroscopic slip boundary condition by introducing a combination coefficient to control the boundary slip, which includes:

The combined specular with diffusive reflection boundary condition (MR)[65][56][98][99], which can be written as:

$$f_i(x) = (1-\sigma)f_i^{sr}(x) + \sigma f_i^{dr}(x) \quad (28)$$

The combined bounce back with specular reflection boundary condition ( BR )[53], which can be written as:

$$f_i(x) = (1-r)f_i^{sr}(x) + rf_i^{bb}(x) \quad (29)$$

And the combined bounce back with diffusive reflection boundary conditions ( MB )[57], which is:

$$f_i(x) = (1-\chi)f_i^{bb}(x) + \chi f_i^{dr}(x) \quad (30)$$

Historically, there is no consensus on how to choose combination coefficients[54][57][60][61][65][100]. Most Recent studies indicate that the analytic solution of LBM with Bounce back or other slip boundary conditions in Poiseuille flow is just a parabolic profile of N-S equation shifted by a numerical slip, $u_s$, and by setting this numerical slip $u_s$ to Mawellian slip boundary condition, the combination coefficients can be obtained[57][59]. A detailed analysis of three kinds of combined form of boundary conditions are given by Zheng et al.[101] in which they pointed out that if one chooses combination coefficient equals to tangential momentum accommodation coefficient (TMAC)(TMAC=0.8 is used in their case studies), the discrete effects of the boundary conditions will induce large numerical errors. Specifically, MR overestimates the velocity slip, BR and MB underestimate the velocity slip (see Fig. 5), and three boundary conditions cause large errors in the slip velocity ( $> 60\%$ for the BR scheme, $> 20\%$ for the MB scheme, and $> 40\%$ for the MR scheme) within the slip regime (see Fig. 6). For micro-tube flow, to match Maxwell-type second order slip boundary:

$$u_s = A_1 Kn \frac{\partial u}{\partial y}\bigg|_{wall} - A_2 Kn^2 \frac{\partial^2 u}{\partial y^2}\bigg|_{wall} \quad (31)$$

Where $u_s$ is the slip velocity, $A_1$ and $A_2$ are the first order and second order slip coefficients respectively. If one chooses the combination coefficients as the following forms (Eq.32 to Eq.34 ), the three kinds of combined forms are identical in a parametric range, and the discrete effects caused by boundary conditions can be moved out (see RBR, RMB and RMR in Fig.5 and Fig.6 ).

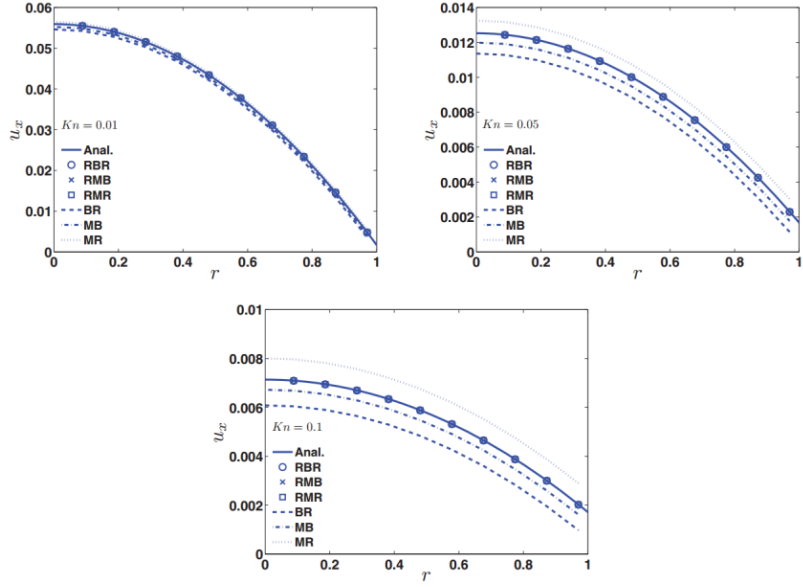

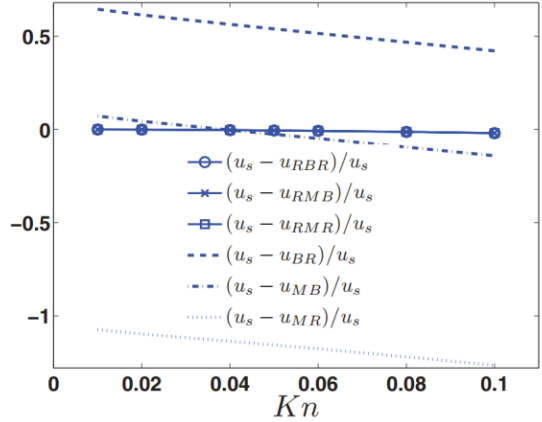

Figure 5: Velocity profiles with TMAC=0.8[101]

Figure 6: Relative error of velocity slip vs Kn with TMAC=0.8[101]

$$\sigma = \left\{1 - \sqrt{\frac{\pi}{6}}\left[A_1 + \left(A_2 + \frac{8}{\pi}\right)Kn\right] + \frac{3\pi^{1/2}\Delta^2}{8\sqrt{6}Kn}\right\}^{-1} \quad (32)$$

$$r = 2 - \left\{\frac{1}{2} - \sqrt{\frac{\pi}{6}}\left[\frac{A_1}{2} + \left(\frac{A_2}{2} + \frac{4}{\pi}\right)Kn\right] + \frac{3\pi^{1/2}\Delta^2}{8\sqrt{6}Kn}\right\}^{-1} \quad (33)$$

$$\chi = \left\{\frac{1}{2} - \sqrt{\frac{\pi}{6}}\left[\frac{A_1}{2} + \left(\frac{A_2}{2} + \frac{4}{\pi}\right)Kn\right] + \frac{3\pi^{1/2}\Delta^2}{8\sqrt{6}Kn}\right\}^{-1} \quad (34)$$

The limitation of the combined forms is that they still have some difficulties in predicting flow as the coefficients have to be specified as inputs from the matching of empirical or analytical descriptions of the simulation. In other words, the boundary slip observed from combined forms are more phenomenal than physical as those from other methods such as DSMC and Boltzmann equation[98][102]. In order to solve this problem and enable the LBM to derive a predictive value, Sbragaglia and Succi[52] suggested a micro-scale simulation, such as MD to obtain the values of coefficients corresponding to

### 2.2.5 Langmuir slip boundary (LSB)

Another slip boundary used in the LBM is based on Langmuir slip model. The Langmuir slip model has been developed by by Eu et al.[103] and Myong[104]. The slip velocity can be expressed as:

$$u_s = \alpha u_w + (1-\alpha)u_g \tag{35}$$

Where, $u_g$ denotes the velocity adjacent to the wall, $\alpha$ is the fraction of surface coved by adsorbed atoms at thermal equilibrium which varies with the type of gas and the nature of wall material. For monatomic gases and diatomic gases $\alpha$ can be expressed by:

$$\alpha|_{mon} = \frac{\beta p}{1+\beta p}; \alpha|_{di} = \frac{\sqrt{\beta p}}{1+\sqrt{\beta p}} \tag{36}$$

with:

$$\beta = \frac{A\lambda/Kn}{k_b T} exp(\frac{D_e}{K_b T}) = \frac{1}{4\omega Kn} \tag{37}$$

Where, $k_b$ represents the Boltzmann constant, $A$ is the mean area of a site, $D_e$ is the potential adsorption parameter and $T$ is the temperature. Compared to Maxwell slip model, in which the slip coefficient is a free parameter from the concept of diffusive reflection, the major advantage of Langmuir slip model is to attribute a physical coefficient of heat and gas particle interaction potential, $\omega$[104], Nevertheless, Langmuir slip model still has the same difficulty in determining the value of the coefficient as the Maxwell slip model[105].

The LB model with Langmuir slip boundary condition was firstly used to study gas bearing problems[106]. Later, Chen and Tian[107] implement Langmuir slip boundary by non-equilibrium extrapolation scheme to study gas flow in the micro-channel, with an approximation that the local gas density at the wall equals to gas density at the nearby cell, the distribution functions for wall boundary nodes can be expressed as:

$$f_i(x_b) \approx \alpha[f_i^{eq}(\rho(x_f), u(x_b)) - f_i^{eq}(\rho(x_f), u(x_f))] + f(\rho(x_f), u(x_f)) \tag{38}$$

Where, $x_b$ is the boundary node, and $x_f$ is the nearest node to the boundary node.

## 2.3 Capture the Knudsen layer effect

For micro-gaseous flow simulation, Once $Kn > 0.1$, presence of Knudsen layer near the solid boundary cannot be ignored. Inside of the Knudsen layer, the intermolecular collisions become insufficient and the quasi thermodynamic-equilibrium assumption, upon which the N-S equation depends, does not hold, and the standard LBM which is only accurate at the N-S level, fails to work(See Fig. 7). Therefore, some lattice Boltzmann schemes or lattice Boltzmann schemes combined with slip boundary conditions have been proposed to capture the flow in side of Knudsen layer and simulate flow with high Kn beyond the slip flow regime which can not be captured with the N-S equation.

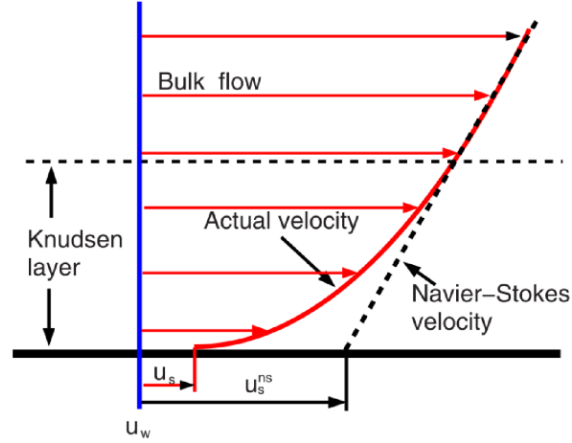

Figure 7: Schematic of the Knudsen layer

### 2.3.1 High order lattice Boltzmann model

Since the LBE is a discrete approximation to the continuous Boltzmann equation which is capable of describing gas flow in a wide range of Knudsen number, some efforts have been made to design higher-order LBMs by increasing discrete velocities to achieve its approximation accuracy to the continuous Boltzmann equation[108-112], It has been shown that the high order LB models can improve the predictions for non-equilibrium flows and the simulation results are qualitatively agree with previous studies[96][109][110], but two problems hinder the application of this approach. Firstly, the accuracy of the higher-order LBM does not increase monotonically with increase of the order of Gauss-Hermite quadrature. Also the approach cannot guarantee an improved accuracy for microscale gas flow with the Knudsen number up to $Kn = O(1)$[113]. Secondly, a priori knowledge on the order of the discrete velocity set should be available for a specific flow, and the use of a high order discrete-velocity set usually means a large computational cost[59][114].

### 2.3.2 Effective viscosity model

The dynamic viscosity $\mu$ of a fluid plays a critical role in its dynamic behaviour and thus is an important consideration for fluid flow simulation. The viscosity describes the internal friction between moving fluid layers and the spanwise momentum conductivity. Furthermore, it determines the relaxation time $\tau$ and $\tau_s$ used in the BGK-LBE model and MRT-LBE model, respectively. With increase in Knudsen number, the gas viscosity in the Knudsen layer gradually deviates from the definition of traditional gas viscosity because of the rarefaction effect, therefore, by building the relationship of viscosity/relaxation time with local $Kn$, and then by taking account of the variation of local $Kn$, the LBM can be extended to simulation high Knudsen number flows.

**Relations of relaxation time with Kn**

By introducing an empirical parameter $a$, Nie et al.[90] firstly built the linear relationship of $Kn$ with relaxation time $\tau = HKn/a + 0.5$. The problem of this model is that the artificial parameter, $a$ needs to be determined by comparing simulation results with that obtained from experiments. Therefore it cannot be transferred to another situation[51]. To improve this, other attempts have been proposed in the literature, and it has been confirmed that the relaxation time can be related to Kn by multiplying a certain microscopic velocity. The choice of this velocity is rather diverse in the literature(see Table 2), and most recently, Zhang et al. and Guo et al.[56][115] proposed that to satisfy a "consistent requirement". this certain microscopic velocity must be chosen as $\sqrt{\pi RT/2}$, and $\tau$ can be given as:

$$\tau = HKn/(\delta x\sqrt{\pi/6}) + 0.5 \tag{39}$$

Table 2: Relationships of relaxation time with Kn

| | |
|---|---|
| Lim et al. [91] | $\tau = HKn/\delta x$ |
| Lee and Lin [50] | $\tau = HKn/\delta x + 0.5$ |
| Niu et al. [51] | $\tau = HKn/(c\sqrt{6/\gamma\pi}) \approx HKn$ |
| Tang et al. [65] | $\tau = (HKn)/(\delta x\sqrt{8/3\pi}) + 0.5$ |
| Zhang et al.[55] Guo et al.[56] | $\tau = HKn/(\delta x\sqrt{\pi/6}) + 0.5$ |

**Wall function approach**

One way to capture the Knudsen layer is to modify the mean free path by implementing a "geometry dependent" correction function to reflect the wall confinement:

$$\lambda_e = \lambda J(r, \lambda) \tag{40}$$

Here, $J(r, \lambda)$ is the correction function which considers the effect of wall surface on the mean free path $\lambda$, and $r$ is distance from the wall. The different correction functions for parallel walls situations and their valid $Kn$ are list in Table 3.

Wall function approach has been widely used to study micro-channel and micro-tube flows. A general problem of wall functions is that it has been derived based on the distance to the wall and therefore it is difficult to deal with complex geometrics and the Knudsen layer overlap effect[62]. To solve this problem, an approximate Stop's expression is proposed by Guo et al.[56] as:

$$J(Kn) = \frac{2}{\pi} arctan(\sqrt{2}Kn^{-3/4}) \tag{41}$$

It can be seen that this correction function is not related to the distance from the wall. In other words, with this formulation it is possible to tackle the problem of Knudsen layer interference in more complex geometry by computing the average of all effective mean free paths.

Table 3: Summarize of different correction functions

|  | $J$ | $Kn$ |
|---|---|---|
| Stops [116] | $1 + \frac{1}{2}\begin{bmatrix} (r/\lambda - 1)exp(-r/\lambda) \\ +((H-r)/\lambda - 1)exp(-(H-r)/\lambda) \\ -(r/\lambda)^2 E_i(r/\lambda) \\ -((H-r)/\lambda)^2 E_i((H-r)/\lambda) \end{bmatrix}$ | $\approx 0.2$ |
| Zhang et al. [88] | $\frac{1}{1+0.7[e^{-Cr/\lambda}+e^{-C(H-r)/\lambda}]}$ | $\approx 1$ |
| Arlemark et al. [117] | $1 - \frac{1}{82}\begin{bmatrix} exp\left(-\frac{H/2+y}{\lambda}\right) + exp\left(-\frac{H/2-r}{\lambda}\right) \\ +4\sum_{i=1}^{7} exp\left(-\frac{H/2+r}{\lambda cos\left[\frac{(2i-1)\pi}{28}\right]}\right) \\ +4\sum_{i=1}^{7} exp\left(-\frac{H/2-r}{\lambda cos\left[\frac{(2i-1)\pi}{28}\right]}\right) \\ +2\sum_{i=1}^{6} exp\left(-\frac{H/2+r}{\lambda cos\left[\frac{i\pi}{14}\right]}\right) \\ +2\sum_{i=1}^{6} exp\left(-\frac{H/2-r}{\lambda cos\left[\frac{i\pi}{14}\right]}\right) \end{bmatrix}$ | $\approx 0.2$ |
| Dongari et al. [118] | $1 - \frac{1}{96}\begin{bmatrix} \left(1+\frac{r}{\lambda}\right)^{-2} + \left(1+\frac{H-r}{\lambda}\right)^{-2} \\ +4\sum_{i=1}^{8}\left(1+\frac{r}{\lambda cos\left[\frac{(2i-1)\pi}{32}\right]}\right)^{-2} \\ +4\sum_{i=1}^{8}\left(1+\frac{H-r}{\lambda cos\left[\frac{(2i-1)\pi}{32}\right]}\right)^{-2} \\ +2\sum_{i=1}^{7}\left(1+\frac{r}{\lambda cos\left[\frac{i\pi}{16}\right]}\right)^{-2} \\ +2\sum_{i=1}^{7}\left(1+\frac{H-r}{\lambda cos\left[\frac{i\pi}{16}\right]}\right)^{-2} \end{bmatrix}$ | $\approx 2$ |

**Bosanquet-type effective viscosity approach**

The larger the Knudsen number gets the more converges the mean free path to the characteristic scale $H$. In the case $\lambda \gg H$, the effective mean free path is $H$ and the viscosity, $\mu_\infty = a_\infty \rho \bar{c} H$ with $a_\infty$ being a numerical constant. Michalis et al.[119] observed that the effective viscosity can be approximated by the terms $\mu$ and $\mu_\infty$. They conducted micro-flow simulations with the Direct Simulation Monte Carlo (DSMC) method to investigate the rarefaction effects on the viscosity in the transition regime. The calculated densities and velocity profiles were used to compute the actual viscosity of the fluid. With these results Michalis et al.[119] showed that the rarefaction effects could be captured satisfactorily by a Bosanquet-type effective viscosity approximation which is expressed as:

$$\mu_e \approx \frac{\mu}{1+aKn} \quad (42)$$

Where, the rarefaction factor, $a = a_0/a_\infty$. Because the Bosanquet-type effective viscosity considers the overall rarefaction effect on gas viscosity such as the approximate wall function approach proposed by Guo et al.[56], it can be utilized in rarefied flow simulations in porous media, however, the choice of $a$ is still an open question. Beskok and Karniadakis[18] used $a = 2.2$ together with their general velocity slip boundary condition for simulation of fluid flow in channel. Later, Sun and Chan[120] used $a = 2$ in estimating effective viscosity at different Knudsen number in channels with aspect ratio between 15 and 20. The authors found the results of effective viscosity reasonable when compared with that obtained from the DSMC method. Michalis et al.[119] observed that value of $a$ depends on the Knudsen number and it varies slightly over the majority of the transition flow regime. Recently, Kalarakis et al.[121] suggested $a = 3.4$ based on a study in which authors estimated permeability using LB method and then compared the results with that from DSMC for porous media with porosity equals to 0.7 and 0.8.

## 2.4 Evaluation of the LBM for micro-gaseous flow simulation

In order to evaluate the LBM in capturing micro gas flow, other numerical simulations, such as DSMC, analytical solutions of Boltzmann equation or N-S equation and experimental observation are used for comparison. Comparative analysis and discussions of the use of the LBM for micro-gaseous flow by different authors are summarised and presented in Table 4,5,6,7. Because most literatures qualitatively claim their scheme to be accurate in certain $Kn$ range, it is hard to conclude the best model for simulation micro gas flow unless some acceptable errors are put forward.

Table 4: Comparisons of the LBM with other methods for isothermal micro-gaseous flow

| | Geometry | Validation model | Comparison parameter | $Kn$ |
|---|---|---|---|---|
| Lim et al 2002 [91] | pressure driven micro-channel flow | first-order analytical solutions of N-S [122] experimental results of UCLA[123] | velocity pressure mass flow rate | 0.056 0.155 |
| | shear driven micro-channel flow | DSMC analytical soluton (not specified) | velocity | 0.01 |
| Niu et al 2004[51] | planar Couette flow | DSMC[124] MD[125] Maxwells prediction[125] | slip length | $< 0.1$ |
| | pressure driven micro-channel flow | experimental results of Pone et al (1994)[123] first-order analytical solutions of N-S[122] | pressure mass flow rate | 0.053 |
| Tang 2004 [54] | pressure driven micro-channel flow | first-order analytical solutions of N-S[122] experimental results[126] | pressure velocity friction constant mass flow rate | 0.055 0.16 |
| Lee and Lin 2005[50] | force driven micro-channel flow | DSMC[127] linearized Boltzmann solution[128] | velocity | $\leqslant 0.1$ |
| | pressure driven micro-channel flow | first-order analytical solutions of N-S[122] | velocity pressure mass flow rate | 0.025 0.05 0.1 |
| Tang et al 2005[65] | micro-Couette flow | first-order analytical solutions of N-S[122] | velocity | 0.05 |
| | micro-Poiseuille flow with constant inlet velocity | first[122]and second-order solution[129] of outlet friction constant | fraction constant | 0.1 |

Table 5: Comparisons of the LBM with other methods for isothermal micro-gaseous flow

| | Geometry | Validation model | Comparison parameter | $Kn$ |
|---|---|---|---|---|
| Zhang et al 2005[55] | micro-channel flow | linearized Boltzmann solution[128] | mass flow rate | $\leqslant 0.4$ |
| Guo et al 2006[56] | pressure driven micro-channel flow | DSMC and IP[130] | velocity pressure | 0.0194 0.194 0.388 |
| Zhang et al 2006[88] | planar Couette flow | DSMC and IP[131] | velocity pressure | 0.01,0.25 0.75,0.5,1.0 |
| | pressure driven micro-channel flow | linearized Boltzmann solution[128] | velocity | 0.113,0.226, 0.451,0.677,1.13 |
| Niu et al 2007[96] | planar Couette flow | DSMC[132] | velocity | 0.1 1 |
| | force driven micro-channel flow | DSMC[132] linearized Boltzmann solution[133] | velcity mass flow rate | up to 10 $Kn_{min}$ at $Kn \approx 1$ |
| Guo et al 2008[59] | planar Couette flow | DSMC provide by Hongwei Liu | velcity | 0.1,0.25,0.5 0.75,1.0,1.5 |
| | force driven micro-channel flow | linearized Boltzmann solution[128] | velcity mass flow rate | 0.1128,0.2257,0.4514 0.6670,0.9027,1.1284 2.2568,4.5135 |
| Tang et al 2008[62] | planar Couette flow | linearized Boltzmann solution [134] | velocity | 0.1 − 10 |
| | force driven micro-channel flow | linearized Boltzmann solution [128] | velocity | 0.1 − 10 |

Table 6: Comparisons of the LBM with other methods for isothermal micro-gaseous flow

| | Geometry | Validation model | Comparison parameter | $Kn$ |
|---|---|---|---|---|
| Li et al 2011[61] | force driven poiseuille flow | linearized Boltzmann solution[128] second order solution of N-S [129] | velocity | 0.1128,0.2257,0.4514 0.6670,0.9027,1.1284 2.2568,4.5135 |
| | | solution of Boltzmann equation [135] | mass flow rate | up to 3 $Kn_{min}$ at $Kn \approx 0.9$ |
| | pressure driven micro-channel flow | DSMC[130] IP-DSMC[130] first-order analytical solutions of N-S[122] | velcity pressure | 0.194 0.388 |
| | | experiment data [136][137] first-order analytical solutions of N-S[122] solutions of second-order slip model[138] | mass flow rate | up to 0.5 $Kn_{min}$ at $Kn \approx 0.9$ |
| Zhuo and Zhong, 2013[60] | force driven poiseuille flow | linearized Boltzmann solution[128] | velocity | 0.1129,0.4514,1.1284 2.2568,4.5135,6.7703 9.0270,11.2838 |
| | | solution of Boltzmann equation[135] | mass flow rate | up to 10 $Kn_{min}$ at $Kn \approx 0.9$ |
| | pressure driven micro-channel flow | DSMC[130] IP-DSMC[130] first-order analytical solutions of N-S[122] | velcity pressure | 0.0194 0.388 |
| | | experiment data[136][137] solutions of second-order slip model[138] | mass flow rate | $Kn_{out}$ up to 0.5 $Kn_{out} \in [0.05, 0.8]$ |

Table 7: Comparisons of the LBM with other methods for isothermal micro-gaseous flow

| | Geometry | Validation model | Comparison parameter | $Kn$ |
|---|---|---|---|---|
| Esfahani and Norouzi, 2014[68] | pressure driven poiseuille flow | DSMC[130] IP-DSMC[130] first-order analytical solutions of N-S[57] | streamwise velocity | 0.0194 0.194 0.388 |
| | | first-order analytical solutions of N-S[57] | normal velocity | 0.0194 0.194 0.388 |
| | | linearized Boltzmann solution[128] solutions of second-order slip model[139] | mass flow rate | up to 10 $Kn_{min}$ at $Kn \approx 0.5$ |

# 3  Simulation of fluid/gas flow in shale gas Reservoirs with the LBM

Shale gas reservoirs are considered unique because of its complex porous structures and petrophysical properties. Generally, theoretical treatments of flow in porous shale are usually associated with three different length scales: pore-(microscopic), representative elementary volume (macroscopic)- and field- scales. Pore scale is the smallest scale where the flow is studied on pore geometries. Pore scale results can provide quantities such as permeability, porosity at various locations. With these results, some fundamental issues such as medium variability can be quantitatively assessed and REV can be quantified. The REV scale is larger than the pore scale but much smaller than the field scale, within the range of an REV, the macroscopic variables (such as permeability and porosity) do not change with the magnitude of the averaging volume. As an image based simulation tool, the LBM has been developed to simulate fluid/gas flow in porous shale on pore scale and REV scale.

## 3.1 Pore scale simulation

At the pore scale, the flow of fluid/gas through the pores of shale reservoir is directly simulated by the LBM. Initially, standard LBM is developed with non-slip boundary conditions to study continuum flow in porous media. As gas flow in shale is involved with different flow regimes, recently, some attempts have been done to extended standard LBM to capture non-continuum phenomenon.

### 3.1.1 Estimation of intrinsic permeability

Under the continuum assumption, standard LBM is an attractive simulation method, and can be easily applied in complex boundary geometries due to its kinetic nature and a simple bounce back rule for non-slip boundary condition. This particular feature makes the LBM superior to classical numerical techniques (e.g., finite differences and finite elements) and other simplified network models for studying flow in realistic porous media. Moreover, as the incompressible N-S equation can be obtained in the nearly incompressible limit of the LB model, LBM is widely used to estimate the intrinsic permeability of shale along with tomography techniques[140][141][142]. For example, Chen et al.[143] generated 3D FIB-SEM images of shale samples and estimated intrinsic permeability and tortuosity with standard LBM. Latterly the same techniques were applied but the computation was carried out using a pore-scale GPU-accelerated LB simulator (GALBS) which increases the computing speed (1000 times faster than the serial code and 10 times faster than the paralleled code run on a standalone CPU)[144]. The relative permeability of shale was also studied by Cantisano et al.[145] and Nagarajan et al.[102]. Nagarajan et al.[146] calculated the gas-oil relative permeability of Liquid Rich Shale (LRS) with the LBM and compared their results with laboratory studies. The authors observed that standard LBM can give similar remaining oil saturation to that of the experimental observations, however, significant difference exists in the relative permeability curves.

### 3.1.2 Estimation of apparent permeability

When continuum hypothesis breaks down, the gas molecules tend to "slip" on the solid surface, and the measured gas permeability through a porous media is higher than that of intrinsic permeability due to gas slippage. To estimate the gas apparent

permeability($k_a$) of shale, several approaches have been proposed from the literature.

**Klinkenberg model based LBM**

In Klinkenberg model, the apparent permeability is calculated based on a linear correlation factor($f_c$) for correcting the intrinsic permeability ($k_0$):
$$k_a = k_0 f_c \quad (43)$$

Where, $f_c$ is a correction factor and is given by[147]:
$$f_c = \left(1 + \frac{b_k}{p}\right) \quad (44)$$

With a slip factor $b_k$ which depends on $Kn$. Later, it is confirmed that Klinkenberg's correlation is only first order accurate. Beskok and Karniadakis[18] proposed a second-order correlation by considering the different flow regimes from continuum flow to free molecular flow:
$$f_c = [1 + \alpha(Kn)Kn]\left[1 + \frac{4Kn}{1 - bKn}\right] \quad (45)$$

Where, slip coefficient $b$ equals to $-1$ for slip flow, and $\alpha(Kn)$ is a rarefaction coefficient which is given by Florence et al.[21] for a purely diffusive (TMAC=1) situation as:
$$\alpha(Kn) = \frac{128}{15\pi^2} tan^{-1}[4Kn^{0.4}] \quad (46)$$

Allan and Mavko[148] firstly used the combination of Beskok and Karniadakis - Florence's correlation along with a 3D incompressible LB model to estimate the $k_a$ for a shale image with a 167nm length in each dimension. In their study, the intrinsic permeability $k_0$ was predicted by standard LBM. Further, the adsorption gas was induced in their model as an immobile phase which affects permeability in two manners: decreasing the gas permeability and changing the TMAC. Moreover, they pointed out that a supercritical phase transition may take place during the pressure depression and diffusive flow mechanisms becomes a negligible mass transport mechanism when gas is in a supercritical phase (see Fig. 8).

Allan and Mavko's work appears to be the first to apply the LBM to quantify the effect of slip and adsorption on micro-porous shale rock transport properties. With an average $Kn$ which is derived from flux weighted average pore width, however, the realistic gas flow through the pore space cannot be obtained. Also, the surface diffusion of adsorption gas is ignored in their model, and no further discussions are provided on the influence of adsorption gas on TMAC.

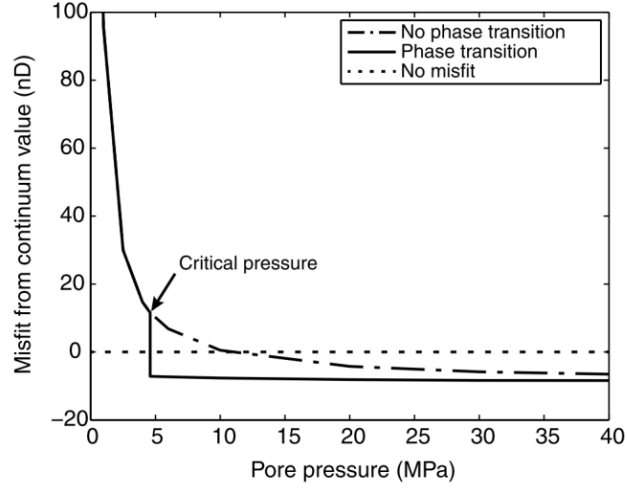

Figure 8: The difference between the total permeability and the continuum permeability as a function of pore pressure. For comparison, two curves are presented corresponding to a gaseous phase (dot-dashed curve), and to a fluid phase that undergoes a supercritical transition (solid curve). The horizontal dotted line marks that the total permeability is equal to the continuum permeability when adsorption layer is not considered.

**Dusty gas model (DGM) based LBM**

The dusty gas model is based on the superposition of convection and molecular spatial diffusion(Knudsen diffusion):

$$J = -\frac{\rho k_a}{\mu} = J_d + J_k = -\frac{\rho k_d}{\mu}\nabla p + -\frac{\rho}{p}D_{eff}\nabla p \qquad (47)$$

Where, $J$ is the mass flux per unit area, $J_d$ is the viscous flow flux and $J_k$ is the Knudsen diffusion flux. $D_{eff}$ is the effective Knudsen diffusivity. Based on Eq.47, the apparent permeability can be calculated as:

$$k_a = k_0(1 + \frac{D_{eff}\mu}{pk_0}) \qquad (48)$$

Very recently, a DGM based LB model was proposed by Chen et al.[149] to calculate the apparent permeability of shale. In which a MRT-LBM for fluid flow and a SRT-LBM for mass transport were used to estimate intrinsic permeability $k_0$ and effective Knudsen diffusivity $D_{eff}$, separately. In order to reflect the variation of local Knudsen diffusivity with local pore diameter, the relaxation time used in the LBM transport model was modified based on:

$$\tau_g = \frac{d_p}{d_{p,ref}}(\tau_{p,ref} - 0.5) + 0.5 \qquad (49)$$

Where $d_p$ is local pore diameter, $d_{p,ref}$ is a reference pore size which is chosen as 25 nm in their simulation, and $\tau_{p,ref}$ is set as 1.0.

The DGM based LB model developed by Chen et al. was adopted to estimate the tortuosity and the gas apparent permeability of four reconstructed shale samples form Sichuan Basin (China). Their simulation results indicate that commonly used Bruggeman equation greatly underestimates tortuosity of shale, and DSM based LB model can give a competitive results of apparent permeability as that given by Beskok and Karniadakis-Civan's correction[150](see Fig.9).

Later, the same techniques were applied by them to estimate the apparent permeability of 250 reconstructed kerogen samples using a reconstructed method called overlapping sphere method, with a porosity ranges from 0.1 to 0.55, and a mean diameter equals to 30, 45, and 60 nm respectively[151]. The simulation results further confirm the high turtuosity of kerogen and Beskok and Karniadakis-Civan's correlation for calculating apparent permeability of shale matrix.

Chen et al's work is the first study of numerically investigating the effective Knudsen diffusivity based on real porous structures of shale with the LBM. The advantage of this method is ease of implementation. In their DGM based LBM, gas adsorption and surface diffusion are ignored, and the Knudsen diffusion term used in DGM contains the assumption of fully diffusive boundary condition (TMAC=1),which underestimates the mass flux in the transition regime[152].

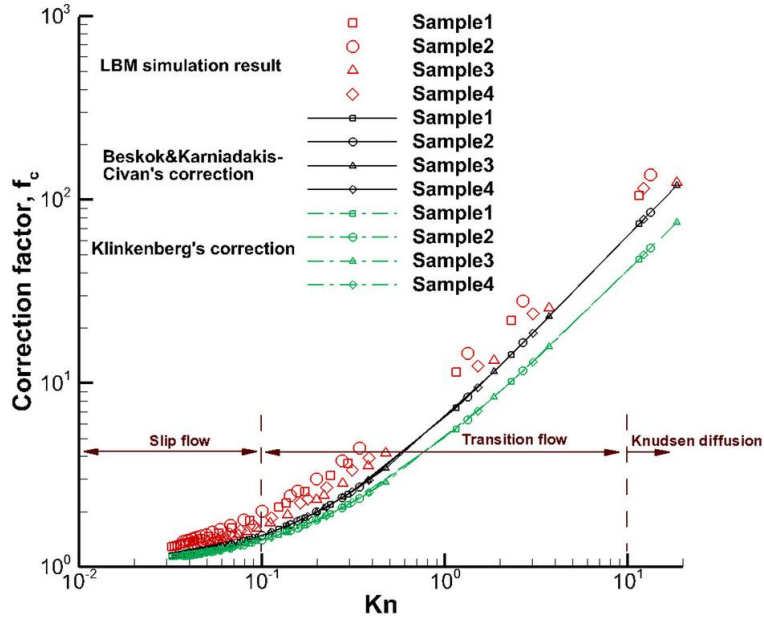

Figure 9: Correction factors predicted by the LBM and empirical correlations under different Knu

**Slip N-S model based LBM**

Besides Klinkenberg model and DSM based LBM, Slip N-S model based LBM is the direct simulation of gas flow through pore structure with the use of appropriate slip boundary conditions and effective relaxation time(See section 2). A few attempts have been made in the literature by applying slip LBM to shale gas flow through kerogen pores. In these applications, additional physical properties are added into slip boundary conditions and LB model to reflect the adsorption gas and/or surface diffusion effect.

Fathi and Akkutlu[153] developed a LB-Langmuir isotherm model in which LSB was used to capture the velocity slip. To consider the impact of surface diffusion, $u_w$ in Eq. 35 was calculated based on:

$$u_w = u_a = \frac{D_s}{D_k \rho_a}\left[\rho_g \mu_g \frac{KC_{\mu s}}{(1+KC)^2}\right] \quad (50)$$

Where, $C$ is the free gas density, $C_{\mu s}$ is the maximum adsorption capacity, $D_k$ is the tortuosity-corrected coefficient of molecular diffusion, $D_s$ is surface diffusion coefficient, $\rho_a$ is the adsorbed-phase density, and $K$ is the equilibrium partition (or distribution) coefficient. Interestingly, the pseudopotential model proposed by Shan and Chen[154] was also employed in their LB model to consider the non-ideal gas effect and the interactions between solid and gas, however, it is argued that this treatment may leads to a double consideration of the gas-solid interactions[155] as pseudopotential model and slip LB are parallel in capturing gas slippage[85][86][87].

Based on BR and Langmuir isotherm model, Ren et al. proposed a different form of LB slip boundary condition, in which the surface diffusion of adsorption gas is modelled as a moving wall. In their model, the transport rate of adsorbed gas is independent of $D_k$ as mentioned in Eq.50 and can be expressed as:

$$u_a = -D_s \frac{\rho_s M}{\rho_a V_{std}} \frac{q_L p_L}{(p_L + p_{free})^2} \frac{\partial p_{free}}{\partial x} \quad (51)$$

Where, $M$ is the molecular weight of the gas, $\rho_s$ is the organic solid density, $V_{std}$ is the gas volume per mole at standard temperature and pressure, $q_L$, $p_L$ and $p_{free}$ are Langmuir volume, Langmuir pressure, and free-gas pressure, separately. They demonstrated that their model can predict more reasonable physical behaviour compared to that in Fathi and Akkutlu[153], however, no validation is provided when considering the adsorption gas and surface diffusion effect. The non-ideal gas effect was also studied by Ren et al.[156](see Fig.10), and they demonstrated the necessity to use LB under real gas conditions instead of the one under ideal gas conditions as large difference exists between the simulation results for ideal and non-ideal gas.

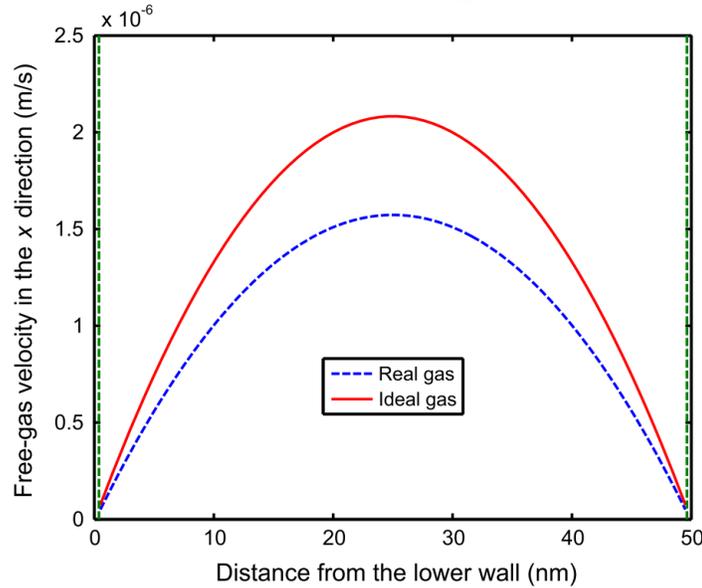

Figure 10: Comparison of free-gas velocity profiles in Kerogen pores using the LB under ideal and non-ideal gas conditions

Slip N-S based LBM gets ride of the usage of empirical or semi-empirical correlations, and it is believed that it can provide accurate simulation results comparable with experimental studies. The main advantage of this method is that detailed local information of the flow(such as velocity field and pressure field) can be obtained and which can be used to study macroscopic relations, however, Slip N-S based LBM is still in its infancy, previous applications in shale are limited to single channel flow. Current complex boundary treatments prevent its further application in porous media[85][97].

### 3.2 REV scale simulation

Due to the huge computational cost, pore-scale LBM is impractical to perform REV-scale simulations of porous shale. Therefore, several alternative approaches are proposed to apply the LBM on REV scale[35][157][158].

To apply the LBM for REV scale flow simulation, Guo et al.[49] proposed a generalized LB model, in which the forcing term( Eq. 10 ) was related to the porosity of porous media:

$$F_i = (1 - \frac{1}{2\tau})\omega_i[\frac{c_i-u}{c_s^2} + \frac{c_i \cdot u}{\varepsilon c_s^4}c_i] \cdot F \tag{52}$$

With

$$F = -\frac{\varepsilon\mu}{k}u - \frac{\varepsilon\rho F_\varepsilon}{\sqrt{k}}|u|u + \varepsilon\rho G \tag{53}$$

Where, $F_\varepsilon$ is the geometry function and $k$ is the intrinsic permeability $k_0$. Both $F_\varepsilon$ and $k$ need to be estimated based on the empirical relationships with porosity $\varepsilon$.

Chen et al.[159] further extended this generalized LBM for slip flow by considering Klinkenberg's effect on REV scale, and this was achieved by the usage of apparent permeability $k_a$ instead of $k_0$. Beskok and Karniadakis-Civan's correlation[19] was adopted to calculate $k_a$:

$$k_a = k_0 \left[1 + \frac{1.358}{1+0.17Kn^{-0.4348}}Kn\right]\left[1 + \frac{4Kn}{1+Kn}\right] \tag{54}$$

In this equation, $k_0$ is calculated based on Kozeny-Carman (KC) equation[160], and the local characteristic pore radius for the calculation of $Kn$ was estimated by[161]:

$$r = 0.08886\sqrt{\frac{k_0}{\varepsilon}} \tag{55}$$

Chen et al. performed several simulations based on a heterogeneous shale matrix with natural fractures, organic matter and inorganic minerals. Their simulation results qualitatively and quantitatively confirm the increasing of permeability induced by Klinkenberg's effect. Moreover, Chen et al's study provides a framework of applying the LBM on REV scale for further research, within this framework, other physical effects (such as adsorption and surface diffusion) can be easily integrated by modifying the local apparent permeability.

## 4 Discussions and Conclusions

The LB method has gone through significant improvements over the years and has become a viable and efficient substitute for N-S solver in many flow problems. Because of the underlying kinetic nature, LB equation has attracted a huge interest in its extension to mimic micro-gaseous flows. In this paper we presented a general review of the LBM with an emphasis on boundary conditions, treatments for relaxation time and application of the LBM in isothermal fluid/gas flow simulation in shale gas reservoir.

It has been found that the LBM is an effective method for simulation micro gas flow in continuum to slip flow regime. For example, through the Chapman-Enskog expansion, Lattice Boltzmann equation covers the macroscopic continuity and momentum (Navier-Stokes) equations. In slip flow regime, the Knudsen layer takes a small portion of the channel height, and it can be neglected by extrapolating the bulk gas flow towards the wall. In this case, with the implementation of proper relaxation time and combination coefficient for various boundaries such as BB, SR or MD, LB equation can give similar results to that of N-S equation with slip boundary conditions for pressure driven/force driven micro-channel flow and micro-Couette flow.

In transition flow regime, the Knudsen layer effect is significant, and the N-S equation with first-order slip boundary breaks down. Second-order or high-order slip boundary conditions are need, and the change of the local mean free path/Knudsen number inside of the Knudsen layer has to be considered. The validation with MD or DSMC simulation results indicates that the LBM can be extended to simulate gas flow in transition flow regimes with high order velocity sites and/or using wall function approach or approximated viscosity approach, however, because of current effective viscosity models are only accurate with a moderate range of Knudsen number and the degree of freedom in the momentum space is very limited, difficulties still exist in studying non-equilibrium gas flow with high Knudsen numbers using LBM.

Several LBM approaches are proposed to study gas transport in shale gas reservoir both on pore scale and REV scale. Pore scale simulation is an effective way to improve the understanding of different flow mechanisms. Application of the LBM to study shale gas transport on pore scale depends very much on its ability to capture micro-gaseous flow and the transport of adsorption gas. Slip N-S based LBM has the potential to provide accurate simulation results comparable with experimental studies with small Knudsen number. Previous slip N-S based LB models are premature to accurately estimate micro-flow properties in porous shale, and Klinkenberg model or DGM based LBM provide an alternative way to simulate gas flow on pore scale. The development of REV scale based LBM enable us to simulate gas flow on a larger scale, however, the simulation study should be complemented by laboratory studies on core samples. Increase in efficiency of these methods for simulation of micro-gaseous flow have the potential for applications in evaluating shale gas reservoirs.

The LBM still has a number of limitations in studying micro-gaseous flow, such as capturing the gas-solid interactions and non-ideal gas effect. Compared to other methods such as MD and DSMC, previous applications of the LBM are more phenomenal than physical. For the simulation of gas flow in shale gas reservoir, studies based on experiments, MD and DSMC are recommended to be carried out before using the LBM.

# 5 Acknowledgement

The authors would like to acknowledge the support from LANL and School of Petroleum Engineering, UNSW. The second author thanks the support from National Nature Science Foundation of China(No. 51406145 and 51136004). The first author would also like to acknowledge the support from China Scholarship Council(CSC).